\documentclass[conference, a4paper]{IEEEtran}
\IEEEoverridecommandlockouts
\usepackage{caption}
\usepackage[caption=false,font=footnotesize]{subfig}
\usepackage{graphicx}
\usepackage{tabularx}
\usepackage{amsmath,amssymb}
\usepackage{cite}
\usepackage{pgfplots}
\usepackage[utf8]{inputenc}
\usepackage{import}
\usepackage{pgf,tikz}
\usepackage{balance}
\usepackage{enumerate}

\ifCLASSINFOpdf
\else
\fi
\author{\IEEEauthorblockN{Quentin Bodinier,
		Faouzi Bader, and
		Jacques Palicot} 
	\IEEEauthorblockA{SCEE/IETR - CentraleSupélec, Rennes, France, \\}
	\IEEEauthorblockA{Email : \{firstname.lastname\}@supelec.fr}\vspace{-10pt}}

\begin{document}

	\DeclareGraphicsExtensions{eps}
	\graphicspath{{fig/}}
	\bstctlcite{IEEEexample:BSTcontrol}
	\title{%Why Filter Bank Waveforms Will not Drastically Facilitate Coexistence with Legacy CP-
		Coexistence of Filter Banks and CP-OFDM:\\What are the Real Gains?}
	\IEEEspecialpapernotice{(Invited Paper)}
	\maketitle
	
	\begin{abstract}
		A coexistence scenario between filter bank (FB) based waveforms and legacy Cyclic Prefix-Orthogonal Frequency Division Multiplexing (CP-OFDM) is studied. It is shown that studies available on that matter use a wrong metric to rate the interference between the coexisting systems. Specifically, it is shown that even well spectrally localized FB waveforms interfere at a high level with  incumbent CP-OFDM receivers. The presented results invalidate a number of studies in the literature, which stated that FB based waveforms could be very efficiently used to insert communications in parts of spectrum left free by incumbent legacy CP-OFDM users.
		%Indeed, the poor frequency localization and the high asynchronism sensitivity of the latter is not fit to 

		%.  an exact model of interference between OFDM/OQAM and CP-OFDM users is still lacking. In this paper, we fill this gap by deriving exact closed forms of cross-interference between OFDM/OQAM and CP-OFDM users. Our results show that using OFDM/OQAM only marginally reduces interference to legacy users, in contradiction with many results in the literature.
		
		%In the fifth generation of wireless networks (5G), it is expected that secondary users could reuse some parts of the spectrum left free by incumbent Cyclic Prefix-Orthogonal Frequency Division Multiplexing (CP-OFDM) based Long Term Evolution-Advanced (LTE-A) users to optimize the use of the spectrum.
		%A number of papers have stated that using the newly proposed  filteOFDM/Offset-Quadrature Amplitude Modulation (OFDM/OQAM) in conjunction with the PHYDYASr would highly facilitate this operation by limiting interference to the incumbent users. However, recent work has shown that these results were based on a flawn model. To study these coexistence scenarios with more accuracy, we derive in this work the exact closed form of cross-interference that gets injected between an incumbent CP-OFDM and a secondary OFDM/OQAM user.	
	\end{abstract}
	

	% \fi
	\IEEEpeerreviewmaketitle

	\section{Introduction}
	\label{sec:intro}
	In the course of the development of the 5th generation of wireless networks (5G), a number of new waveforms have been proposed to replace Cyclic-Prefix Orthogonal Frequency Division Multiplexing (CP-OFDM) used in Long Term Evolution-Advanced (LTE-A)\cite{Banelli2014}. Indeed, the latter suffers from well-known limitations. Due to the rectangular shaping of symbols in the time domain, CP-OFDM has high out of band emissions (OOBE) and is very sensitive to asynchronism \cite{Mahmoud2009, Medjahdi2010a}.

These two characteristics are very problematic for certain applications and use cases that are expected to be enabled with the advent of 5G. Especially, communication types are expected to become ever more various and diverse, ranging from classical cellular networking to Device-To-Device (D2D) or Machine-To-Machine (M2M) communication \cite{Forschung2013c}. This new paradigm requires the physical layer (PHY) of 5G to be adaptable to various situations, and robust to asynchronous interference coming from neighboring communication devices~\cite{Banelli2014}. 

Besides, these new communication types will add a new burden to the radio spectrum, which is already saturated. To answer this challenge, two directions have mainly been explored: 
\begin{enumerate}[(i)]
	\item Exploit new parts of the spectrum at higher frequencies above $6$ GHz.
	\item Exploit parts of the already licensed spectrum that are temporally left free by incumbent users.
\end{enumerate}
In this paper, we focus on the point (ii). The main challenge to overcome when re-exploiting some parts of the already licensed spectrum lies in the fact that secondary users should insert their communication without causing harmful interference to the incumbent legacy users. In the context of 5G, reusable spectrum mainly belongs to either LTE-A cellular networks, Wi-Fi or TV bands. In all these cases, inserted secondary users will have to coexist with CP-OFDM based incumbent communications, as it is the PHY used by these three technologies.

To protect the incumbent CP-OFDM users, the common policy states that secondary users should respect a certain spectrum mask at the transmitter side. This spectrum mask is usually specified by the standard followed by the incumbent users. CP-OFDM, suffering from poor spectral containment, can hardly fit efficiently in these spectrum masks and is therefore not efficiently applicable to secondary users \cite{Mahmoud2009, Medjahdi2010a,Ihalainen2008,Datta2011a}. 
This observation has motivated the research community to look for new waveforms whose OOBE are decreased compared to CP-OFDM, so that the secondary transmissions could fit into the specified spectrum masks.

New waveform schemes proposed for such applications all fall into the category of filter bank (FB) based multi-carrier waveforms \cite{Saltzberg1967,Farhang-Boroujeny2011}. They all rely on the filtering of the transmit signal by highly frequency selective filters to reduce OOBE. Therefore, users based on this type of waveform easily fit the requirements set by incumbent users in terms of Power Spectral Density (PSD). Building on this, a number of papers have investigated the benefits of using FB based waveforms for the insertion of secondary users \cite{Noguet2011,shaat2010computationally,skrzypczak2012ofdm,Su2016}. Some real-world demonstrations of coexistence between FB waveforms and CP-OFDM have also been undertaken by both industrials and academics \cite{KeysightWaveforms, EurecomExp, Berg2014743}. 

All these studies rely on the observation of the PSD of the waveform used by the secondary. The advantages of using FB waveforms instead of CP-OFDM for secondary users are then rated in terms of the maximum performances that each waveform can achieve under the same spectral mask. However, PSD only accounts for the properties of the transmit signal, and totally neglects the effects of the CP-OFDM incumbent reception. To refine this measurement, it is necessary to study the Error Vector Magnitude (EVM) of the CP-OFDM incumbent receiver when it suffers interference from a neighboring FB waveform. To the best of our knowledge, this is only done experimentally by Keysight Technologies in \cite{KeysightWaveforms} but no comparison of the EVM of the CP-OFDM incumbent as a function of the secondary waveform is given. In \cite{EurecomExp, Berg2014743}, the authors show that FB based secondary users do interfere less than CP-OFDM secondary onto CP-OFDM incumbent receivers. However, the gains shown are not in line with the much higher theoretical gains expected for example in \cite{Noguet2011,shaat2010computationally,skrzypczak2012ofdm}.

In previous works, we already showed that results based on the PSD were misguiding, as they underestimate the interference that is seen after demodulation of the incoming signal by the CP-OFDM incumbent receiver \cite{Bodinier2016ICT}. We further showed that, considering EVM measurements, the interference caused by secondary users is not drastically reduced if the latter use FB waveforms instead of CP-OFDM \cite{Bodinier2016ICT}. As these previous results are contradictory with many results published in the literature, it is important to justify them clearly and revise some previously published results with regards to these new findings.

Therefore, in this paper, we further explain why coexistence of FB waveforms with CP-OFDM is not feasible. To do so, we show that FB waveforms belong to a class of waveforms that are not orthogonal with CP-OFDM. We propose a simplified model to replace the current misguiding PSD-based spectrum masks. Furthermore, we select a representative set of results from the literature and update them according to our interference measurements based on EVM. Both our justifications and the revised set of results we present confirm that coexistence with CP-OFDM is not drastically improved if the secondary users use FB waveforms. 

The remainder of this paper is organized as follows: In Section \ref{sec:model}, we present a canonical model of coexistence. Besides, we explain the difference between the PSD based approach of the literature and the EVM based measurements of interference. In Section \ref{sec:analysis}, we show that proposed FB waveforms are not orthogonal with the CP-OFDM receiver, which causes high interference to the latter. In Section \ref{sec:results} we select a specific set of results available in the literature on coexistence and update them according to our results. Finally, Section \ref{sec:ccl} concludes this paper.

	\section{In-band Coexistence Scenario}
	\label{sec:model}
	\subsection{Typical Coexistence Scenario}
\begin{figure}[!t]
	\captionsetup{justification=justified}
	\subfloat[]{	\begin{tikzpicture}[xscale=0.95]
		\node (I) [blue, text width=2.25cm, align = center, draw] at (0,-0.25) {Incumbent CP-OFDM Transmitting User};
		\node (S) [orange, text width=2.25cm, align = center, draw] at (0,1.75) {Secondary FB based Transmitting User};
		\draw (2,0.75) circle(0.4) ;
		\draw (2,1.1) -- (2,0.4);
		\draw (1.65,0.75) -- (2.35,0.75);
		\draw [>=latex,->] (S.east)-- (2, 1.5) node[above]{$\mathbf{s}_\textbf{s}$}-- (2, 1.15) ;
		\draw [>=latex,->] (I.east)-- (2, 0) node[below]{$\mathbf{s}_\textbf{i}$} -- (2, 0.35) ;
		\node[blue, draw, dashed, text width = 4cm, text height= 2.5cm ] (RX) at (5, 0.75){};
		\node[blue,above] at (RX.north){Incumbent CP-OFDM Receiving User};
		\node[draw, blue, align=center, text width = 1.2cm, align = center] (CPR) at (3.8,0.75) {CP\\ Removal};
		\node[draw, blue, align=center, text width = 1.2cm, align = center] (FFT) at (5.8,0.75) {OFDM Demod.};
		\draw[->,>=latex] (2.4,0.75) -- node[near start,above,pos=0.3]{$\mathbf{r}$} (3.1,0.75);
		\draw[->,>=latex] (CPR)-- node[above,pos=0.5]{$\mathbf{\tilde{r}}$}(FFT);
		\draw[->,>=latex] (FFT)-- node[above,pos=0.5]{$\mathbf{\hat{d}}$} (7,0.75);
		\end{tikzpicture}\label{fig:coex_layout}}\\
	\subfloat[]{	\begin{tikzpicture}[xscale=0.95]
		\draw[->,>=latex] (0,0) -- node[at end, below left]{$\frac{f}{\Delta \text{F}}$} (\linewidth,0);
		\draw[xstep=0.3,ystep=100,very thin] (0.3,0) grid (\linewidth*0.85,0.2);
		\node[below] at  (\linewidth*0.85, 0) {$\frac{M}{2}-1$};
		\node[below] at  (0.2, 0) {$-\frac{M}{2}$};
		\draw[->,>=latex] (\linewidth*0.439, -0.3) -- node[at start,left]{$0$} (\linewidth*0.439, 1.5);
		\draw[dashed,blue] (0.75, 0) -- (0.75, 1.5);
		\draw[dashed,blue] (3.15, 0) -- (3.15, 1.5);
		\draw[dashed,orange] (4.05, 0) -- (4.05, 1.5);
		\draw[dashed,orange] (6.15, 0) -- (6.15, 1.5);
		\node[blue] at (1.9,1.2) {$\mathcal{M}_\textbf{i}$}; 
		\node[orange] at (5.1,1.2) {$\mathcal{M}_\textbf{s}$};
		\draw[xstep=0.3, shift={(-0.15,0)},blue](0.9,0) grid (3.3,1);
		\draw[xstep=0.3, shift={(-0.15,0)},orange](4.2,0) grid (6.3,1);
		%\draw[->,->=latex] (4.05,1) -- node[near end, above]{$\delta_\text{f}$} (4.2,1);
		\end{tikzpicture}\label{fig:coex_spec_view}}
	
	\caption{Considered coexistence scenario\\
		(a) Coexistence layout : the signals of the secondary and the incumbent sum up and are passed through the demodulator of the CP-OFDM incumbent receiver.\\
		(b) Spectral representation : the incumbent and secondary systems coexist in the same spectral band, and each one is assigned a different subset of subcarriers. %The secondary system misaligns its subcarriers with respect to the frequency basis of the incumbent by a frequency offset $\delta_\text{f}$.
	}
	
\end{figure}
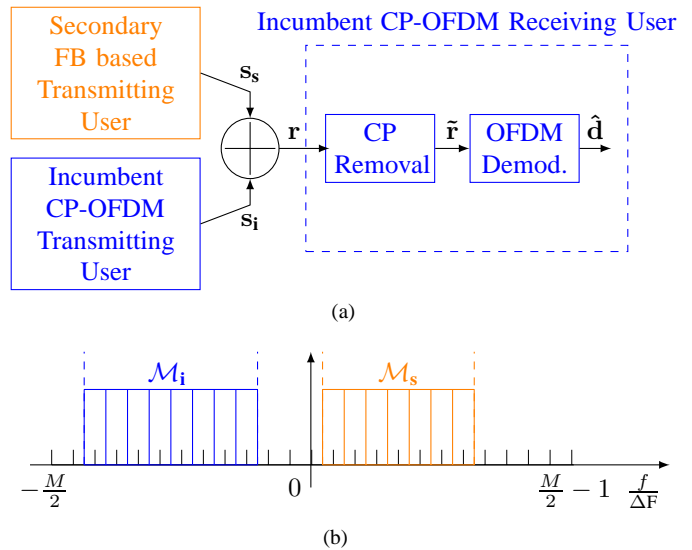

A typical coexistence scenario is presented in Fig.~\ref{fig:coex_layout}, where we see that the signals of both the FB based secondary $\mathbf{s}_\textbf{s}$ and the CP-OFDM based incumbent $\mathbf{s}_\textbf{i}$ add up to form the signal $\mathbf{r}$ at the input antenna of the incumbent CP-OFDM receiver. Note that, here, no noise and no channel effects are considered, in an effort to focus on the interference caused by the FB secondary tranmission. Then, the over-the-air signal $\mathbf{r}$ is passed through the CP-removal block which outputs the useful part of the signal, noted $\mathbf{\tilde{r}}$. After demodulation, the vector of estimated symbols  $\mathbf{\hat{d}}$ is obtained. Under this simplified coexistence scenario, $\mathbf{\hat{d}} = \mathbf{d}+\pmb{\eta}$, where $\mathbf{d}$ is the vector of quadrature amplitude modulation (QAM) symbols modulated by the CP-OFDM incumbent transmitter and $\pmb{\eta}$ is the interference caused by the  secondary FB based user.
In Fig.~\ref{fig:coex_spec_view}, we present an example spectral distribution of the incumbent and secondary users. The two users coexist in the same spectral band composed of $M$ subcarriers of width $\Delta\text{F}$, and each one is assigned a different set of subcarriers, $\mathcal{M}_\textbf{s}$ for the secondary and $\mathcal{M}_\textbf{i}$ for the incumbent.

Under this model, the interference power caused by the secondary FB user onto the incumbent CP-OFDM receiver is expressed as the EVM seen after demodulation as
\begin{equation}
I_\text{tot} = E\{|\mathbf{\hat{d}}-\mathbf{d}|^2\} = E\{|\pmb{\eta}|^2\},
\label{eq:I1}
\end{equation}
where $E\{.\}$ is the mathematical expectation. $I_\text{tot}$ can be further decomposed in the sum of the power of interference caused by each subcarrier of the secondary onto each subcarrier of the incumbent as follows \cite{Bodinier2016ICT}:
\begin{equation}
I_\text{tot} = \sum_{m_\textbf{i}\in\mathcal{M}_\textbf{i},m_\textbf{s}\in\mathcal{M}_\textbf{s}}I(l = m_\textbf{s}-m_\textbf{i}), 
\label{eq:I_tot}
\end{equation}
where  $I(l = m_\textbf{s}-m_\textbf{i})$ is the interference caused by the subcarrier $m_\textbf{s}$ of the secondary to the subcarrier $m_\textbf{i}$ of the incumbent such that the spectral distance between these two subcarriers, noted $l$, is equal to $m_\textbf{s} -m_\textbf{i}$. 

$I(l)$ represents the interference power that is injected by a given subcarrier of the secondary FB user onto a subcarrier at spectral distance $l$ of the incumbent CP-OFDM user. It is the key factor that will determine if it is possible to insert efficiently the secondary transmission. Indeed, the faster $I(l)$ decreases with $l$, the more efficiently the secondary user will be able to utilize the free part of the band without interfering onto the subcarriers of the incumbent user. It is therefore crucial to model this value with great accuracy. 

\subsection{Modeling of $I(l)$ in the literature}
In the literature, the modeling of $I(l)$ is based on the PSD of each subcarrier of the secondary FB transmitter. This model has been initially proposed to rate interference between various OFDM users in a spectrum pooling context in \cite{weiss04}. Precisely, naming $\Phi_\mathbf{s}$ the PSD of each subcarrier of the FB based waveform used by the secondary, $I(l)$ is usually modeled according to the PSD-based model as
\begin{equation}
I_\text{PSD}(l) = \int\limits_{l-\frac{\Delta\text{F}}{2}}^{l+\frac{\Delta\text{F}}{2}} \Phi_\mathbf{s}(f)\ df.
\label{eq:I_PSD}
\end{equation}
The total interference is then obtained by putting \eqref{eq:I_PSD} in \eqref{eq:I_tot}. The obtained expression of the total interference seen by the incumbent receiver is then equivalent to
\begin{equation}
I_\text{PSD,tot} = \int\limits_{\mathcal{M}_\textbf{i}}\mathcal{S}_\textbf{s}(f) df,
\label{eq:I_PSD_tot}
\end{equation}
where $\mathcal{S}_\textbf{s}(f)$ is the total PSD of the signal $\mathbf{s}_\textbf{s}$.
\makeatletter
\def\ifabsgreater #1#2{\ifpdfabsnum 
	\dimexpr#1pt>\dimexpr#2pt\relax
	\expandafter\@firstoftwo\else\expandafter\@secondoftwo\fi }
\makeatother

%%\begin{figure}[!t]
%\begin{tikzpicture}
%\begin{axis}[
%axis lines=middle,
%xmin=-1,xmax=16,
%ymin=-80,
%ymax=4,
%domain=0.01:16,
%samples = 1000,
%]
%\addplot[draw=blue] {20*(ln(abs(sin(pi*deg(x))/(pi*x)))/ln(10))};
%\addplot[draw=orange] {20*ln(abs(sin(pi*deg(x))/(pi*x)+0.971960*(sin(pi*deg(x-1/4))/(pi*(x-1/4))+sin(pi*deg(x+1/4))/(pi*(x+1/4)))+1/sqrt(2)*(sin(pi*deg(x-1/2))/(pi*(x-1/2))+sin(pi*deg(x+1/2))/(pi*(x+1/2)))))/ln(10)};
%\end{axis}
%\end{tikzpicture}
%\end{figure}

From the expression \eqref{eq:I_PSD_tot}, it is clear that the PSD-based model considers interference in the channel, before the input antenna of the incumbent CP-OFDM receiver, and is not consistent with the actual expression of the interference expressed in \eqref{eq:I1}.

	\section{Observation of FB Signal after CP-OFDM Demodulation}
	\label{sec:analysis}
	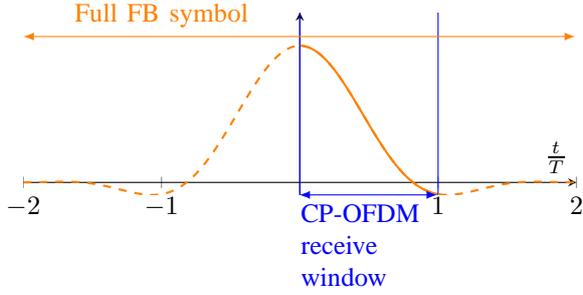
\begin{figure}
		\vspace{-10pt}
	\centering

	\begin{tikzpicture}
	\begin{axis}[xlabel=$\frac{t}{T}$,ylabel= ,ymajorticks = false,domain=-2:2, axis lines=middle,ymax=6, samples=100,smooth,width=\linewidth, height=4cm]
	\addplot[domain = 0:1,mark=none,thick,orange] {1+0.971960*2*cos(deg(2*pi*x/4))+1/sqrt(2)*2*cos(deg(4*pi*x/4))+0.235147*2*cos(deg(6*pi*x/4))};
	\addplot[domain = -2:2,dashed,mark=none,thick,orange] {1+0.971960*2*cos(deg(2*pi*x/4))+1/sqrt(2)*2*cos(deg(4*pi*x/4))+0.235147*2*cos(deg(6*pi*x/4))};
	\draw[blue] (axis cs:1,\pgfkeysvalueof{/pgfplots/ymin}) -- (axis cs:1,\pgfkeysvalueof{/pgfplots/ymax});
	\draw[blue] (axis cs:0,\pgfkeysvalueof{/pgfplots/ymin}) -- (axis cs:0,\pgfkeysvalueof{/pgfplots/ymax});
	\end{axis}
	\draw[<->,>=latex,orange] (0,2.1cm) -- node[above,near start]{Full FB symbol} (\linewidth*.82,2.1cm);
	\draw[<->,>=latex,blue] (\linewidth*.41,0) -- node[below,pos=0.5,text width = \linewidth*0.2]{CP-OFDM receive window} (\linewidth*.615,0);
	\end{tikzpicture}
	\caption{Truncation of the FB filter operated by the CP-OFDM receiver on the receiving window corresponding to a single CP-OFDM symbol.}
	\label{fig:truncation}
		\vspace{-10pt}
\end{figure}
In the former section, we showed that the PSD-based model is not relevant to study the interference $\pmb{\eta}$ and therefore, to rate the EVM $I_\text{tot}$ as defined in \eqref{eq:I1}, which corresponds to the actual value of interference seen by the CP-OFDM incumbent receiver after the OFDM demodulation operations. However, one may think that the PSD-based model is still a good approximation that will give a good idea of the actual values of interference. Hereafter, we explain why it is not the case.

FB waveforms rely on the filtering of each symbol by a filter $g$ that is longer than the time symbol $T=\frac{1}{\Delta\text{F}}$. This means that each FB symbol has a temporal support higher than the time symbol $T$. On the opposite, the CP-OFDM receiver of the incumbent truncates the incoming signal in windows of duration $T$ before performing FFT operations on each of these windows. Therefore, the FB and CP-OFDM signals are not orthogonal, as they are not based on the same temporal support. An example of the way the filter $g$ used by the FB is cut is shown in Fig.~\ref{fig:truncation}. It is shown that only a certain portion of each FB symbol, represented by a solid line, is taken into account by the CP-OFDM receive window. More detailed explanation on this aspect can be found in \cite{Bodinier2016ICT}.

\begin{figure}
	\vspace{-10pt}
	\includegraphics[width=\linewidth]{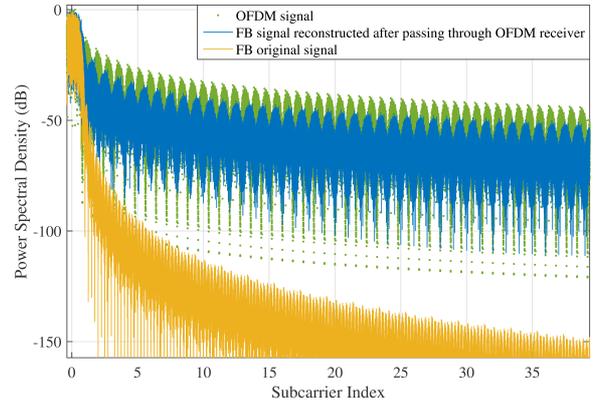}
	\caption{Comparison of the PSD of the FB signal before and after passing through the CP-OFDM demodulator. The CP-OFDM receive operations destroy the good spectral shape of the FB signal, which becomes almost as poorly spectrally localized as an OFDM signal.}
	\label{fig:avant_apres}
	\vspace{-10pt}
\end{figure}
Note that the rigorous mathematical analysis of the effects of this truncation is out of the scope of this paper, but has been led for a particular FB waveform in \cite{Bodinier2016GC}. However, it is readily understood that the rectangular truncation in time of the FB filter  $g$ is equivalent, in frequency, to convolving it with a function $\text{sinc}(fT)$.
From this observation, it is clear that the selective properties of the FB filter $g$ are lost when passing through the CP-OFDM receiver. To illustrate this, we study a particular type of FB waveform, OFDM/OQAM with the PHYDYAS filter. More information on this waveform can be found in \cite{Bellanger2010}. In Fig.~\ref{fig:avant_apres}, we show the PSD of one OFDM/OQAM subcarrier. It is clear from that figure that, after passing through the CP-OFDM receiver, the spectral shape of the OFDM/OQAM signal is deeply altered. Whereas the original OFDM/OQAM signal is almost perfectly localized in frequency, high frequency ripples appear after the CP-OFDM receiver, which are due to the rectangular truncation shown in Fig.~\ref{fig:avant_apres}. Actually, the spectral shape of the altered OFDM/OQAM signal is very close to that of an OFDM signal.
To strengthen our point, we compare in Fig.~\ref{fig:psd_evm} the two ways of computing the values of interference injected by the OFDM/OQAM secondary onto a CP-OFDM incumbent, either through EVM or through PSD. It is very clear from this figure that the PSD measurements are completely irrelevant to measure the interference actually seen by the CP-OFDM incumbent, and do not approximate it in a satisfying way at all.

Though the results presented in Fig.~\ref{fig:avant_apres} and Fig.~\ref{fig:psd_evm} concern here only a particular waveform, the same behavior will be shown by other FB waveforms, as the limiting factor is the receive window of the CP-OFDM incumbent. More generally, because of the effects shown in Fig.~\ref{fig:truncation}, waveforms with different temporal support are bound to cause high interference onto each other. As a matter of fact, the values of $I(l)$ based on EVM measurements have been obtained through simulations in \cite{BodinierICC2016} for a number of other waveforms, such as Generalized Frequency Division Multiplexing (GFDM) \cite{Michailow2014}, Filtered Multi-Tone (FMT) \cite{FMT_Emphatic} and Lapped FBMC \cite{Bellanger2015}. All these waveforms were shown to interfere approximately as much as OFDM/OQAM onto CP-OFDM incumbent users in \cite{BodinierICC2016}.

These results show that, in the context of 5G, to cope with heterogeneity at the PHY level in the network, \textit{usual spectrum masks need to be replaced to efficiently protect incumbent CP-OFDM users}. Indeed, FB based users could easily fit the currently used spectrum masks based on PSD, but still interfere in a critical way onto CP-OFDM incumbent users. More specifically, instead of defining the transmit masks of secondary users, future standards should include the possibility for each incumbent user to define its maximum acceptable EVM post-demodulation, and let the secondary users adapt their transmission to respect that limitation. These new types of masks will be more complex to specify than PSD based spectrum masks, as they need both the properties of the transmitter and the receiver to be taken into account. 

\begin{figure}
	\vspace{-10pt}
	\includegraphics[width=\linewidth]{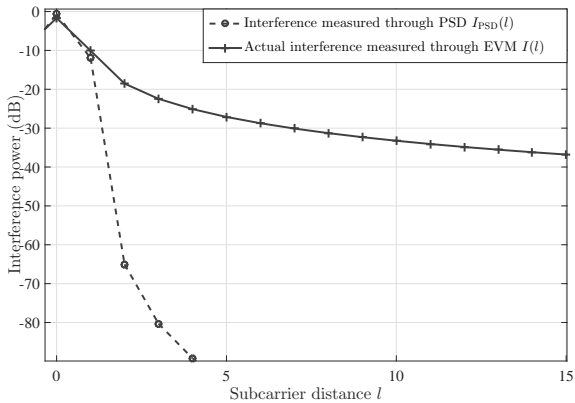}
	\caption{Comparison of values of interference caused by FB secondary onto CP-OFDM primary when using either PSD or EVM measurements.}
	\label{fig:psd_evm}
	\vspace{-10pt}
\end{figure}

	\section{Revising Published Results}
	\label{sec:results}
	In this section, we present two coexistence scenarios that are usually studied in the literature according to the PSD-based model, which we invalidated in the previous sections. We compare results shown in the literature with the ones we obtain when rating the interference caused onto the CP-OFDM incumbent through measurements of EVM after demodulation. Once again, we show results for OFDM/OQAM with Phydyas filter only, but the nature of the presented results is extensible to other FB based waveforms.

\subsection{Reduction of Guard Bands}
A first way of studying the impact of the waveform used by the secondary user on the amount of interference injected onto the incumbent is to consider the width of the guard band that is necessary to protect the incumbent system. This is the approach followed for example in \cite{Ihalainen2008, Datta2011a,skrzypczak2012ofdm,KeysightWaveforms,Berg2014743}. 

\begin{figure}
	\vspace{-10pt}
	\includegraphics[width=\linewidth]{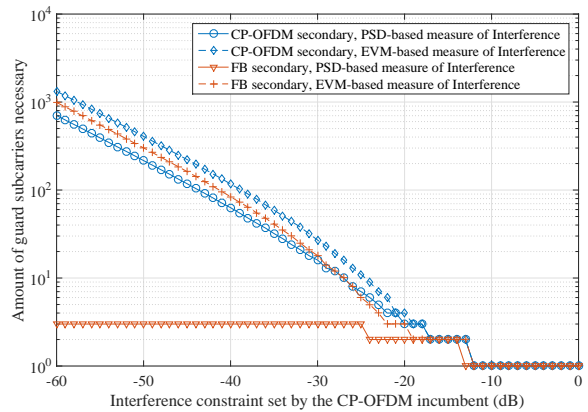}
	\caption{Number of guard subcarriers necessary between the secondary and the incumbent as a function of the interference constraint set by the incumbent.}
	\label{fig:guard bands}
	\vspace{-10pt}
\end{figure}

Here we study this problem in the following layout: we consider that the secondary and the incumbent users both use $20$ subcarriers, and we measure the number of guard subcarriers  that is necessary between them to reach the amount of protection expected by the incumbent. Simulation results are reported in Fig.~\ref{fig:guard bands}. According to the PSD based modeling of interference, a maximum of three subcarriers is needed to protect the CP-OFDM incumbent if the secondary utilizes the considered FB waveform. In reality, when the interference is rated through EVM measurements, the gains of using a FB waveform are much more limited. For example, if the CP-OFDM incumbent sets a maximum interference constraint of $-50$ dB, the guard band must be $407$ subcarriers wide for a CP-OFDM secondary and $301$ subcarriers wide for a FB secondary. This is an interesting gain of $26\%$ but is much less than what is expected based on the PSD, which predicts a relative gain of $1-\frac{3}{216}=99\%$.

\subsection{Optimal Power Allocation for Secondary Users}
A second way to study the coexistence scenarios analyzed in this paper in the literature is to follow the approach of \cite{shaat2010computationally}. In that work, the secondary user aims to optimize its capacity while respecting its total power budget and the maximum interference that is allowed by the incumbent.

Here, we update the results presented in \cite{shaat2010computationally}. We consider a coexistence scenario where $60$ subcarriers are available, an incumbent occupies subcarriers $20$ to $39$ and a secondary tries to exploit subcarriers $0$ to $19$ and $40$ to $59$. The secondary uses the PI-algorithm presented in \cite{shaat2010computationally} to distribute its total power of $1W$ on its subcarriers. We show in Fig.~\ref{fig:max_capa} the capacity it achieves as a function of the interference that is tolerated by the incumbent CP-OFDM user. We compare the results obtained when interference is computed through PSD or EVM measurements. In Fig.~\ref{fig:max_capa}, the gains of using an FB waveform at the secondary seem significant when the PSD-based measurements of interference are used as in \cite{shaat2010computationally}, but vanish when the actual EVM measurements of interference are considered. Note that differences between EVM-based and PSD-based results in the case where both the secondary and the incumbent users utilize CP-OFDM are well explained in \cite{Medjahdi2010a}.

\begin{figure}
	\vspace{-10pt}
	\includegraphics[width=\linewidth]{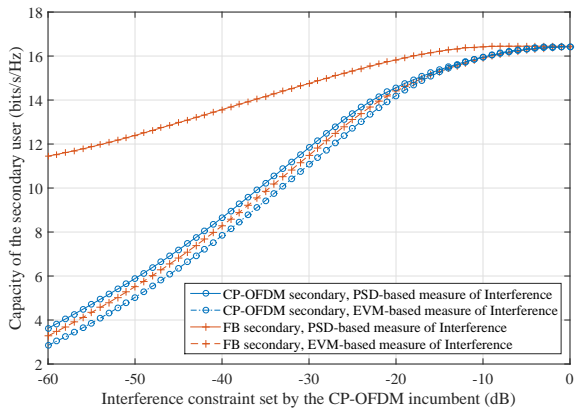}
	\caption{Maximum capacity achieved by the secondary user with the PI-algorithm \cite{shaat2010computationally}. Gains obtained with the use of FB waveforms is significantly decreased when actual EVM values of interference are used instead of the flawed PSD-based approach.}
	\label{fig:max_capa}
	\vspace{-10pt}
\end{figure}
	
	\section{Conclusion}
	\label{sec:ccl}
	In this paper, we explained that the metric commonly used to rate the interference between FB waveforms and CP-OFDM incumbent users does not correspond to the actual interference experienced by the CP-OFDM receiver after the demodulation of the incoming signal. We showed that because the FB signals are not orthogonal with the receiver of the CP-OFDM incumbent receiver, the spectral properties of the former are lost. We demonstrated that the actual gains of using FB based waveforms for coexistence with CP-OFDM are very limited. Based on our observations, we proposed to define new masks based on EVM instead of PSD as it is currently done in the literature in order to efficiently protect the incumbent users. Furthermore, we updated results available in the literature that were built on the PSD-based model and therefore were predicting inaccurate results.

Future work will therefore be twofold. On the one hand, we will develop our analysis of the studied coexistence scenario to find new ways of protecting incumbent CP-OFDM users. The impact of FB interference on synchronization and channel estimation operations performed at the CP-OFDM incumbent receiver should also be considered. On the other hand, we will build a demonstrator to experimentally validate the observations made in this paper and convince the community that FB waveforms proposed so far fail to  protect incumbent CP-OFDM users as much as is expected in the literature.
	
	\section*{Acknowledgement}
	\small This work was partially funded through French National Research Agency (ANR) project ACCENT5 with grant agreement code: ANR-14-CE28-0026-02. The authors also extend their sincere appreciation to Dr. Musbah Shaat, Associate Researcher at CTTC research center (Spain), for his valuable support and help with the PI-algorithm which enabled us to present comparative results in this paper.

	\normalsize
	
%	\section*{Acknowledgment}
%	\small
%	This work was partially funded through French National Research Agency (ANR) project ACCENT5 with grant agreement code: ANR-14-CE28-0026-02.	
	
	
	%\bibliographystyle{IEEEtran}
	\normalsize
	\balance
	\bibliographystyle{IEEEtran}
	\bibliography{IEEEabrv,library}
	
\end{document}